\documentstyle[12pt]{article} 
\def\beq{\begin{equation}}
\def\eeq{\end{equation}}

\def\al{\alpha}
\def\bt{\beta}

\def\De{\Delta}

\def\te{\theta}

\def\lam{\lambda}

\def\ep{\epsilon}

\def\l{\left (}
\def\r{\right )}
\def\fr{\frac}
\def\la{\label}
\def\hs{\hspace}
\def\vs{\vspace}

\def\ran{\rangle}
\def\lan{\langle}
\def\ov{\overline}

\def\tm{\times}

\begin{document}

\begin{titlepage}
\begin{flushright}
BA-02-31\\
\end{flushright}

\begin{center}
{\Large\bf    
Democratic Approach To Atmospheric \\And Solar Neutrino
Oscillations}
\end{center}
\vspace{0.5cm}
\begin{center}
{\large 
{}~Qaisar Shafi$^{a}$
\footnote {E-mail address: shafi@bartol.udel.edu},~ 
{}~Zurab Tavartkiladze$^{b, c}$
\footnote {E-mail address: Z.Tavartkiladze@ThPhys.Uni-Heidelberg.DE} 
}
\vspace{0.5cm}

$^a${\em Bartol Research Institute, University of Delaware,
Newark, DE 19716, USA \\

$^b$ Institute for Theoretical Physics, Heidelberg University, 
Philosophenweg 16, \\
D-69120 Heidelberg, Germany\\

$^c$ Institute of Physics,
Georgian Academy of Sciences, Tbilisi 380077, Georgia\\
}

\end{center}
\vspace{1.0cm}

\begin{abstract}

Working with a ${\cal U}(1)$ flavor symmetry, we show how the hierarchical
structure in the charged fermion sector and a democratic approach for
neutrinos that yields large solar and atmospheric neutrino mixings
can be simultaneously realized in the MSSM framework. However, 
in $SU(5)$ due to the unified multiplets we encounter
difficulties. Namely, democracy for the neutrinos leads to a wrong
hierarchical pattern for charged fermion masses and mixings. 
We discuss how this is overcome in flipped $SU(5)$.

\end{abstract}

\end{titlepage}

\section{Introduction}

Recent SuperKamiokande data appear to confirm the existence of both 
atmospheric \cite{atm} and
solar \cite{sol} neutrino oscillations. From the atmospheric data 
the preferred oscillation parameters are
\beq
\sin^2 2\te_{\mu \tau }\simeq 1~,~~~~~
\De m^2_{\rm atm}\simeq 3\cdot 10^{-3}~{\rm eV}^2~,
\la{atm}
\eeq
while for solar neutrinos the preferred oscillation scenario is a
large angle MSW solution with
\beq
\sin^2 2\te_{e \mu, \tau }\approx 0.8~,~~~~
\De m^2_{\rm sol}\sim 10^{-4}~{\rm eV}^2~.
\la{sol}
\eeq

In attempting to simultaneously accommodate the atmospheric and solar
neutrino
data, one should provide a reasonable theoretical background for
understanding the origin of large (in one case even maximal!) mixings
in (\ref{atm}) and (\ref{sol}). At the same
time, the origin of hierarchies between
charged fermion masses and their CKM mixing angles must be explained.
Finally, one also must find an explanation of how the third mixing angle
$\te_{13}$ in the neutrino sector appears to be 
small($\stackrel{<}{_\sim }0.2$) \cite{chooz}. For a
unified description of
quark-lepton sector, one well motivated idea is that of flavor
symmetries, with an abelian ${\cal U}(1)$ being the simplest possibility.
A variety of models for obtaining the desirable fermion
mass
pattern with ${\cal U}(1)$ have been considered
\cite{feru1}, \cite{1feru1}. The ${\cal U}(1)$ symmetry
also can be promising in the neutrino sector 
\cite{nuu1}-\cite{2maxnu1}, 
especially for generating nearly maximal mixings between
the flavors \cite{maxnu1}-\cite{2maxnu1}. While the atmospheric
neutrino data
strongly suggests maximal mixing, 
for the
solar neutrinos there is significant deviation  from maximal value 
($\sin^2 2\te_{e \mu, \tau }\approx 0.8$).
Because of this, textures leading to bi-maximal neutrino mixings 
\cite{maxnu1}-\cite{2maxnu1}
need to be modified appropriately. This is not always easy  
in the presence of a flavor symmetry such as  ${\cal U}(1)$, 
and one should look
for alternative ways for building up the neutrino sector. One
alternative
(to the maximal mixing texture) is the so called {\it democratic
approach} \cite{dem}, in which lepton doublets of different families have
the same ${\cal U}(1)$ charge. That is, the ${\cal U}(1)$ symmetry does 
not
distinguish them from each other and one could naturally expect large
neutrino
mixings. By the same token, however, the masses of all neutrinos might be
of similar
magnitude, which would be problematic for obtaining the distinct mass
scales relevant for atmospheric and solar neutrinos. This
is easily avoided , however, through a careful choice of the singlets
(right
handed neutrino sector) \cite{1N}, \cite{2maxnu1}.

In contrast with the left handed lepton doublets, the remaining lepton and
quark
superfields should have distinct transformation
properties under ${\cal U}(1)$ in order to obtain desirable hierarchies
between their masses and mixings. Following this strategy, we start our
considerations with MSSM and show that the democratic approach works out
neatly, because MSSM does not provide stringent constraints on the 
${\cal U}(1)$
charge assignments. However, for GUTs the situation can be drastically
changed. Namely, we demonstrate that for $SU(5)$ GUT [with ${\cal U}(1)$
flavor symmetry], the democratic approach gives an unacceptably small
Cabibbo
angle. The root of this problem lies in the unified multiplets and
therefore can be shared by other GUTs unless some 
additional elements are introduced. While in $SU(5)$ it 
may be difficult to realize the democratic approach in a natural way, 
we consider a flipped $SU(5)$ scheme in which the democratic approach for
large neutrino mixings is nicely consistent with the hierarchies in the
charged
fermion sector. We conclude with a brief remark about the third neutrino
mixing angle $\te_{13}$.

\section{${\cal U}(1)$ Flavor Symmetry: Fermion Masses \\And
Neutrino Oscillations}

Let us start our considerations with the MSSM augmented with ${\cal U}(1)$
flavor
symmetry. In addition, we introduce a singlet superfield $X$ with ${\cal
U}(1)$ charge $Q(X)=-1$ and assume that its scalar component has a VEV
\beq
\fr{\lan X\ran }{M_{\rm Pl}}\equiv \ep \simeq 0.2~.
\la{Xvev}
\eeq
$\ep $ plays the role of an expansion parameter and is crucial for the
explanation of hierarchies among the charged fermion masses and their
mixings. With the following assignment of ${\cal U}(1)$ charges for the
quark-lepton superfields
$$
Q[q^{(1)}]=3~,~~Q[q^{(2)}]=2~,~~Q[q^{(3)}]=0~,~~
Q[{u^c}^{(1)}]=4~,~~Q[{u^c}^{(2)}]=1~,~~Q[{u^c}^{(3)}]=0~,
$$
\beq
Q[{d^c}^{(1)}]=n+2~,~~~~Q[{d^c}^{(2)}]=Q[{d^c}^{(3)}]=n   
\la{charges}
\eeq
$$
Q[l^{(1)}]=n-n_3+n_2+n_1~,~~Q[l^{(2)}]=n-n_3+n_2~,~~
Q[l^{(3)}]=n-n_3~,
$$
\beq
Q[{e^c}^{(1)}]=n_3-n_2-n_1+5~,~~Q[{e^c}^{(2)}]=n_3-n_2+2~,~~
Q[{e^c}^{(3)}]=n_3~,~~
\la{leptcharge}
\eeq
($n$, $n_{1, 2, 3}$ are some integers and superscripts stand for
generation indices)
and the pair of higgs doublets $Q(h_u)=Q(h_d)=0$, the relevant couplings
generating the up, down quark and charged lepton masses respectively are
\begin{equation}
\begin{array}{ccc}
 & {\begin{array}{ccc}
\hspace{-5mm} u^c_1 & \,\,~~u^c_2 ~ & \,\,u^c_3 ~
\end{array}}\\ \vspace{2mm}
\begin{array}{c}
q_1 \\ q_2 \\q_3
 \end{array}\!\!\!\!\! &{\left(\begin{array}{ccc}
\,\,\epsilon^7~~ &\,\,\epsilon^4~~ &
\,\,\epsilon^3  
\\  
\,\,\epsilon^6~~   &\,\,\epsilon^3~~  &
\,\,\ep^2
 \\
\,\,\epsilon^4~~ &\,\,\epsilon ~~ &\,\,1
\end{array}\right)h_u }~, 
\end{array}  \!\!  ~~~~~
\begin{array}{ccc}
 & {\begin{array}{ccc}
\hspace{-5mm} ~~d^c_1~ & \,\,d^c_2 ~~ & \,\,d^c_3 ~~~~~~

\end{array}}\\ \vspace{2mm}
\begin{array}{c}  
q_1 \\ q_2 \\q_3
 \end{array}\!\!\!\!\! &{\left(\begin{array}{ccc} 
\,\,\epsilon^5~~ &\,\,\epsilon^3~~ &
\,\,\epsilon^3
\\
\,\,\epsilon^4~~   &\,\,\epsilon^2~~  &
\,\,\epsilon^2
 \\
\,\,\epsilon^2~~ &\,\,1~~ &\,\,1
\end{array}\right)\epsilon^nh_d }~,
\end{array}  \!\!  ~~~~~
\label{updown}
\eeq
\begin{equation}
\begin{array}{ccc}
 & {\begin{array}{ccc}
\hspace{-7mm} e^c_1~~~~~~ & \,\,e^c_2 ~~~ &~~~~ \,\,e^c_3 
  
\end{array}}\\ \vspace{2mm}
\begin{array}{c}
l_1 \\ l_2 \\l_3
 \end{array}\!\!\!\!\! &{\left(\begin{array}{ccc}
\,\,\epsilon^5~~ &\,\,\epsilon^{n_1+2}~~ &
\,\,\epsilon^{n_1+n_2}
\\  
\,\,\epsilon^{5-n_1}~~   &\,\,\epsilon^2~~  &
\,\,\ep^{n_2}
 \\
\,\,\epsilon^{5-n_1-n_2}~~ &\,\,\epsilon^{2-n_2}~~ &\,\,1
\end{array}\right)\epsilon^nh_d }~.
\end{array}  \!\!  ~~~~~
\label{lept}
\eeq
Note that the entries in textures such as (\ref{updown}) and (\ref{lept})
are real and accompanied by factors of order unity. We
will not be concerned with CP violating phases in this work.
Upon diagonalization of (\ref{updown}), (\ref{lept}), for the Yukawa
couplings we obtain
\beq
\lambda_t\sim 1~,~~
\lambda_u :\lambda_c :\lambda_t \sim
\epsilon^7:\epsilon^3 :1~,
\la{ulam}
\eeq
\beq
\lam_b\sim \lam_{\tau}\sim \ep^n ~,~~~
\lambda_d :\lambda_s :\lambda_b \sim
\epsilon^5:\epsilon^2 :1~,
\label{dlam}
\eeq
\beq
\lambda_e :\lambda_{\mu } :\lambda_{\tau } \sim
\epsilon^5:\epsilon^2 :1~,
\label{elam}
\eeq
while for the CKM matrix elements:
\beq
V_{us}\sim \epsilon~,~~~V_{cb}\sim \epsilon^2~,~~~
V_{ub}\sim \epsilon^3~.
\label{ckm}
\eeq
Thus, the ${\cal U}(1)$ flavor symmetry nicely explains the hierarchies
between the
charged fermion masses and CKM mixing angles.

As far as the lepton mixing matrix is concerned, from
(\ref{leptcharge}) 
and the form of (\ref{lept}), one expects\footnote{We do not expect
possible enhancements from the right handed neutrino sector, because it
would need either specific arrangement or some fine tunings.} 
$\sin^2 2\te_{\mu
\tau }\sim \fr{4\ep^{2n_2}}{(1+\ep^{2n_2})^2}$
and $\sin^2 2\te_{e \mu, \tau }\sim \fr{4\ep^{2n_1}}{(1+\ep^{2n_1})^2}$. 
With $n_1=n_2=0$
which means $Q[l^{(1)}]=Q[l^{(2)}]=Q[l^{(3)}]$, one expects
$\sin^2 2\te_{\mu \tau }\sim 1$, $\sin^2 2\te_{e \mu, \tau }\sim 1$. 

To realize oscillations
we have to
generate neutrino masses. Introducing an MSSM singlet neutrino
${\cal N}$  with 
${\cal U}(1)$ charge $Q({\cal N})=p$, with couplings
\beq
\ep^{n+p}(l_1+l_2+l_3){\cal N}h_u+\ep^{2p}M_{\cal N}{\cal N}^2~,
\la{dirmaj}
\eeq
(we assume all entries of order unity) and integrating out ${\cal N}$
leads to a massive state 
$m_{\nu_3}\sim \fr{\ep^{2n}h_u^2}{M_{\cal N}}$. For
${M_{\cal N}}/\ep^{2n}\sim 10^{14}$~GeV, $m_{\nu_3}\sim 0.1$~eV,
which is relevant for atmospheric neutrinos. 
Including a second singlet state ${\cal N}'$ with charge 
$Q({\cal N}')=q$ and couplings
$\ep^{n+q}(l_1+l_2+l_3){\cal N}'h_u+\ep^{2q}{M_{\cal N}}'{{\cal N}'}^2$,
taking ${M_{\cal N}}'/\ep^{2n}\sim 3\cdot 10^{15}$~GeV, and integrating
out 
${\cal N}'$ will introduce into the neutrino mass matrix the deviations 
$\sim \fr{\ep^{2n}h_u^2}{{M_{\cal N}}'}\sim 3\cdot 10^{-3}$~eV. This 
will create a similar order mass for the second light neutrino state. This
mass scale guarantees the large angle MSW oscillations of solar neutrinos.
Thus, with this setting the desirable neutrino mass scales can be obtained
\cite{1N}, \cite{2maxnu1}. Note that large lepton mixings are obtained
due to the same ${\cal U}(1)$ charge
assignments for the left handed lepton doublets, possible in MSSM 
because there were no constraints on $n_{1, 2, 3}$ and $n$ in
(\ref{leptcharge}). 

One would naturally wish to extend this mechanism to SUSY
GUTs. However, it turns out that due to unified multiplets it  is not
a straighforward
task. For example, in $SU(5)$ GUT each family of quark-lepton
superfields is embedded in an anomaly free $10+\bar 5$ superfields, where
$10=(q, u^c, e^c)$ and $\bar 5=(l, d^c)$. Therefore
$Q[q^{(\al )}]=Q[{u^c}^{(\al )}]=Q[{e^c}^{(\al )}]$ and 
$Q[l^{(\al )}]=Q[{d^c}^{(\al )}]$
($\al $ is a
generation index). With universal ${\cal U}(1)$ charges for $l^{(\al )}$
states, one also has the same charges for ${d^c}^{(\al )}$
superfields. For
obtaining the
desirable  hierarchies in (\ref{elam})  for charged leptons, one has to take
$Q[{e^c}^{(3)}]=0$, $Q[{e^c}^{(2)}]=2$, $Q[{e^c}^{(1)}]=5$. 
But this means that
$Q[q^{(3)}]=0$, $Q[q^{(2)}]=2$, $Q[q^{(1)}]=5$. Although this gives a good
estimate for 
$V_{cb}$($\sim \ep^2$), the expected value of Cabibbo angle is 
$\sim \ep^3$, which  is smaller by factor $\sim 25$ than the measured
value
($\sin \te_c\simeq 0.2$).
Thus, in the framework of minimal SUSY
$SU(5)$,
it seems difficult to realize the democratic approach 
discussed above. The reason is the unified multiplets which provide
constraints on the ${\cal U}(1)$ charge
assignments of the MSSM chiral superfields. Of course, one can think of a
possible extension such that the light 
$q^{(\al )}$ and ${e^c}^{(\al )}$ states originate
from different unified multiplets. By introducing some additional
states it might be possible to realize this. However, it is  hard
to imagine such a splitting among leptonic and colored states. Note that
this situation  closely resembles the doublet-triplet (DT) splitting 
problem in the scalar sector and whose resolution in SUSY 
$SU(5)$ requires a rather complicated extensions \cite{DTsu5}.
However, there are GUTs in which DT splitting is acheaved in an elegant
way and flipped $SU(5)$ GUT is one example
\cite{flipsu5}. From
experience in obtaining a natural DT splitting in the scalar sector of 
$SU(5)\tm U(1)$ through the missing partner mechanism, with introduction
of
additional vector-like matter we can
manage to split the unified matter multiplets in such a way that the
democratic
approach to neutrino mixings nicely works out. In the next
section we present the flipped $SU(5)\tm U(1)$ model and its extension.

\section{Flipped $SU(5)$ GUT}

The 'matter' sector of minimal flipped $SU(5)\tm U(1)$ GUT consists of
anomaly
free $\bar 5_3+10_{-1}+1_{-5}$ supermultiplets per generation, where the
subscripts denote $U(1)$ charges and
\beq
\bar 5_{3}=(l, u^c)~,~~~
10_{-1}=(q, d^c, \nu^c)~,~~~
1_{-5}=e^c~.
\la{minfer}
\eeq
The 'higgs' sector contains the following supermultiplets
\beq
H\sim 10_{-1}~,~~\ov{H}\sim \ov{10}_1~,~~\phi \sim 5_2~,~~
\ov{\phi}\sim \bar 5_{-2}~.
\la{scalars}
\eeq
$H$, $\ov{H}$ are responsible for $SU(5)\tm U(1)$ breaking  to 
$SU(3)_c\tm SU(2)_L\tm U(1)_Y\equiv G_{321}$.
$\phi $ and $\ov{\phi }$
contain the MSSM doublet-antidoublet pair $h_d$ and $h_u$ respectively.

Let us first show that the $SU(5)\tm U(1)$ model, supplemented with ${\cal 
U}(1)$ flavor symmetry and with minimal fermion content (\ref{minfer})
neither
yilds the desirable hierarchies between charged fermion masses and
mixings,
nor the two large neutrino mixings.
For the CKM mixing angles we need the hierarchies in (\ref{ckm}). Taking
into account
(\ref{minfer}) we conclude that
\beq
Q[10_{-1}^{(1)}]=3~,~~~
Q[10_{-1}^{(2)}]=2~,~~~
Q[10_{-1}^{(3)}]=0~.
\la{10charge}
\eeq
The down quark masses emerge
from $10_{-1}^{(\al )}10_{-1}^{(\bt )}\phi $  couplings, and with
$Q(\phi )=Q(\ov{\phi })=0$ and (\ref{10charge}) we have
\begin{equation}
\begin{array}{ccc}
&  {\begin{array}{ccc}
\hspace{-5mm}~~10_{-1}^{(1)} & \,\,10_{-1}^{(2)} & \,\,10_{-1}^{(3)} 

\end{array}}\\ \vspace{2mm}
\begin{array}{c}
10_{-1}^{(1)}  \\ 10_{-1}^{(2)}  \\10_{-1}^{(3)} 
 \end{array}\!\!\!\!\! &{\left(\begin{array}{ccc}
\,\,\epsilon^6~~ &\,\,\epsilon^5~~ &
\,\,\epsilon^3
\\
\,\,\epsilon^5~~   &\,\,\epsilon^4~~  &
\,\,\epsilon^2
 \\
\,\,\epsilon^3~~ &\,\,\epsilon^2~~ &\,\,1
\end{array}\right)\phi }~,
\end{array}  \!\!  ~~~~~
\label{dflip}
\end{equation}
which gives the unacceptable ratio $\fr{m_s}{m_b}\sim \ep^4$
(a reasonable value for the latter would be $\sim \ep^2$). 

Morever,
the observed hierarchies for up quark masses in (\ref{ulam}) (generated
through
$10_{-1}^{(\al)}\bar 5_3^{(\bt )}\ov{\phi }$ couplings)
dictates the following assignment
\beq
Q[\bar 5_3^{(1)}]=4~,~~~
Q[\bar 5_3^{(2)}]=1~,~~~
Q[\bar 5_3^{(3)}]=0~.
\la{5charge}
\eeq
Since the $l$ states also come from $\bar 5_3$-plets [see (\ref{minfer})],
according to (\ref{5charge}) we will have 
$Q[l^{(1)}]=4$, $Q[l^{(2)}]=1$, $Q[l^{(3)}]=0$. For the lepton
mixing
elements this gives $V_{23}^l\sim \ep $ and $V_{12}^l\sim \ep^3$, both
of which are in contradiction with observations. We therefore
conclude that the matter sector of flipped 
$SU(5)$ 
model must be extended if ${\cal U}(1)$ flavor
symmetry is invoked.

\subsection{Extended Flipped $SU(5)$}

In the fermion sector we
introduce  three families of vector like states 
$(F+\ov{F})^{(\al )}$ ($\al =1, 2, 3$), where
\beq
F\sim 5_2~,~~~~~~\ov{F}\sim \bar 5_{-2}~.
\la{Fs}
\eeq
In terms of $G_{321}$ they decompose as
\beq
F(5_2)=(l,~\ov{d}^c)_F~,~~~~
\ov{F}(\bar 5_{-2})=(\bar l,~d^c)_{\ov{F}}~.
\la{dec}
\eeq
With these states and including specific couplings one can arrange that
the physical light $l$ and $d^c$ states will come from multiplets
different
from $\bar 5_3$ and $10_{-1}$ respectively. This is realized through 
a way resembling the missing partner mechanism operative in the
higgs sector of
$SU(5)\tm U(1)$. Let us show this in a  one
generation example first. Generalization to three families will be
straightforward. With couplings
\beq
H\bar 5_3\ov{F}+H10_{-1}F+M_F\ov{F}F~,
\la{FH}
\eeq
and assuming that $\lan H\ran \gg M_F$, one can easily verify that
$l_{\bar 5_3}$  and $\bar l_{\ov{F}}$ form a state with
mass
$\sim \lan H\ran \sim M_G$. Therefore, the light left handed doublet state
resides in $F$. At the same time, $d^c_{10_{-1}}$ and
$\ov{d}^c_F$ end up getting mass $\sim \lan H\ran$, and therefore the 
light $d^c$
state comes from $\ov{F}$. This gives us the possibility to build
a realistic
fermion sector with two large neutrino mixings.

Let us then turn to the realistic case of
three
generations. The ${\cal U}(1)$ charge prescriptions for $10_{-1}^{(\al )}$
and 
$\bar 5_3^{(\al )}$  remain the same as in (\ref{10charge}) and
(\ref{5charge}) respectively. For the other states let us make
the assignments
$$
Q[1_{-5}^{(1)}]=5~,~~Q[1_{-5}^{(2)}]=2~,~~Q[1_{-5}^{(3)}]=0~,~~~
Q[F^{(1)}]=Q[F^{(2)}]=Q[F^{(3)}]=0~,    
$$
\beq
Q[\ov{F}^{(1)}]=2~,~~~Q[\ov{F}^{(2)}]=Q[\ov{F}^{(3)}]=0~.
\la{allcharge}
\eeq
{}From (\ref{10charge}), (\ref{5charge}), (\ref{allcharge}) the
couplings responsible for the decoupling of appropriate
states are schematically
\begin{equation}
\begin{array}{ccc}
 & {\begin{array}{ccc}
~\ov{F}^{(1)}&\ov{F}^{(2)}&\,\,\ov{F}^{(3)}~~~
\end{array}}\\ \vspace{2mm}
\begin{array}{c}
\bar 5_3^{(1)}\\ \bar 5_3^{(2)}\\\bar 5_3^{(3)}

\end{array}\!\!\!\!\! &{\left(\begin{array}{ccc}
\,\, \ep^6~~&
\,\,  \ep^4~~ &\,\, \ep^4
\\
\,\, \ep^3 ~~ &\,\,\ep ~~&\,\, \ep ~
\\
\,\, \ep^2 ~~ &\,\,1~~&\,\, 1~
\end{array}\right)H }~,
\end{array}  \!\!~
\begin{array}{ccc}
& {\begin{array}{ccc}
F^{(1)}&F^{(2)}
&F^{(3)}~~
\end{array}}\\ \vspace{2mm}
\begin{array}{c}
10_{-1}^{(1)} \\ 10_{-1}^{(2)} \\
10_{-1}^{(3)} 

\end{array}\!\!\!\!\! &{\left(\begin{array}{ccc}
 \ep^3~~
 &~\ep^3&~~~ \ep^3
\\
 \ep^2~~
&~\ep^2&~~~\ep^2
\\
1~~&~1&~~~1
\end{array}\right)H~,
}
\end{array}
\label{3HF}
\end{equation}                
\begin{equation}
\begin{array}{ccc}
 & {\begin{array}{ccc}
 ~~~\ov{F}^{(1)} & \,\hspace{-1mm}\ov{F}^{(2)}  
&\hspace{-1mm} \ov{F}^{(3)}~~~~~~ 
\end{array}}\\ \vspace{2mm}
\begin{array}{c}
F^{(1)} \\ F^{(2)} \\ F^{(3)}
 \end{array}\!\!\!\!\! &{\left(\begin{array}{ccc}
\,\,\epsilon^2~~ &\,\,1~~ &
\,\,~1
\\
\,\,\epsilon^2~~   &\,\,1~~  &
\,\,~1
 \\
\,\,\epsilon^2~~ &\,\,1~~ &\,\,~1
\end{array}\right)M_F }~.
\end{array}  \!\!  ~~~~~
\label{3FF}
\end{equation}
Let us assume now that $M_F\ll \lan H\ran \ep^4$. {}From the couplings
in (\ref{3HF}), (\ref{3FF}) we realize that the light $l^{(\al )}$ and
${d^c}^{(\al )}$ states respectively come from $F^{(\al )}$ and
$\ov{F}^{(\al )}$
\beq
F^{(\al )} \supset l^{(\al )}~,~~~~
\ov{F}^{(\al )}\supset {d^c}^{(\al )}~.
\la{states}
\eeq
With prescription (\ref{allcharge}) all $F^{(\al )}$ states have the same
${\cal U}(1)$ charges, and according to (\ref{states}) the light left
handed
lepton doublets also have identical (democratic) transformation properties
under ${\cal U}(1)$. The latter guarantee the two neutrino mixings we
are after. At
the same time, the charged fermion masses and mixings have desirable
hierarchies. Namely, the relevant couplings generating up, down quark and
charged lepton masses respectively are 
\begin{equation}
\begin{array}{ccc}
&  {\begin{array}{ccc}
\hspace{-5mm}~~\bar 5_3^{(1)} & \,\bar 5_3^{(2)} & \,\bar 5_3^{(3)} 

\end{array}}\\ \vspace{2mm}
\begin{array}{c}
10_{-1}^{(1)}  \\ 10_{-1}^{(2)}  \\ 10_{-1}^{(3)} 
 \end{array}\!\!\!\!\! &{\left(\begin{array}{ccc}
\,\,\epsilon^7~~ &\,\,\epsilon^4~~ &
\,\,\epsilon^3
\\
\,\,\epsilon^6~~   &\,\,\epsilon^3~~  &
\,\,\epsilon^2
 \\
\,\,\epsilon^4~~ &\,\,\epsilon ~~ &\,\,1
\end{array}\right)\ov{\phi }}~,
\end{array}  \!\!  ~~~~~
\label{upex}
\end{equation}
\begin{equation}
\begin{array}{ccc}
&  {\begin{array}{ccc}
\hspace{-5mm}~~\ov{F}^{(1)} & \,\ov{F}^{(2)} & \,\ov{F}^{(3)}~~

\end{array}}\\ \vspace{2mm}
\begin{array}{c}
10_{-1}^{(1)}  \\ 10_{-1}^{(2)}  \\ 10_{-1}^{(3)} 
 \end{array}\!\!\!\!\! &{\left(\begin{array}{ccc}
\,\,\epsilon^5~~ &\,\,\epsilon^3~~ &
\,\,\epsilon^3
\\
\,\,\epsilon^4~~   &\,\,\epsilon^2~~  &
\,\,\epsilon^2
 \\
\,\,\epsilon^2~~ &\,\,1 ~~ &\,\,1
\end{array}\right)\fr{\ov{H}}{M}\phi }~,
\end{array}  \!\!  ~~~~~
\label{downex}
\end{equation}
\begin{equation}
\begin{array}{ccc}
&  {\begin{array}{ccc}
\hspace{-5mm}~~{F}^{(1)} & \,{F}^{(2)} & \,{F}^{(3)}~~ 

\end{array}}\\ \vspace{2mm}
\begin{array}{c}
1_{-5}^{(1)}  \\ 1_{-5}^{(2)}  \\ 1_{-5}^{(3)} 
 \end{array}\!\!\!\!\! &{\left(\begin{array}{ccc}
\,\,\epsilon^5~~ &\,\,\epsilon^5~~ &
\,\,\epsilon^5
\\
\,\,\epsilon^2~~   &\,\,\epsilon^2~~  &
\,\,\epsilon^2
 \\
\,\,1~~ &\,\, 1 ~~ &\,\,1
\end{array}\right)\fr{\ov{H}}{M}\phi }~,
\end{array}  \!\!  ~~~~~
\label{eex}
\end{equation}
where $M$($\stackrel{>}{_\sim }M_G$) is some cut off scale.
Substituting appropriate VEVs in (\ref{upex}) and upon diagonalization 
we find the  hierarchies in (\ref{ulam}),
while diagonalization of (\ref{downex}),
(\ref{eex}) yield the hierarchies in (\ref{dlam}), (\ref{elam}). Note that
(\ref{upex}), (\ref{downex}) also give rise to  the CKM
mixing 
angles in (\ref{ckm}). At the same time, from
(\ref{eex}), one expects
\beq
\sin^2 2\te_{e \mu , \tau}\sim 1~,~~~~~
\sin^2 2\te_{\mu \tau }\sim 1~.
\la{oscangle}
\eeq
Dirac and Majorana couplings $\nu^clh_u$ and $M_R\nu^c \nu^c$ respectively
are generated through
$10_{-1}F\ov{H}~\ov{\phi }$ and $(10_{-1}\ov{H})^2$ type couplings. In our
scenario all $l^{(\al )}$ ($F^{(\al )}$) states have the same ${\cal
U}(1)$
charges, and to avoid the same mass scales for atmospheric and solar
neutrinos, we will decouple ${\nu^c}^{(1)}$, ${\nu^c}^{(2)}$ states
(from $10_{-1}^{(1, 2)}$). Introducing two singlets $N_{1, 2}$ with
charges $Q(N_{1, 2})=-3, -2$, through the couplings
$(10_{-1}^{(1)}N_1+10_{-1}^{(2)}N_2)\ov{H}$ after substituting $\ov{H}$'s
VEV, the states ${\nu^c}^{(1, 2)}$ decouple with $N_{1, 2}$, and at this
stage
$\nu_{1, 2}$ are massless. {}From the couplings
\beq
\fr{1}{M}10_{-1}^{(3)}F^{(\al )}\ov{H}~\ov{\phi }+
M_R10_{-1}^{(3)}10_{-1}^{(3)}\l \fr{\ov{H}}{M}\r^2~,
\la{1dirmaj}
\eeq
$\nu_3$ obtains a mass $m_{\nu_3}\sim \fr{h_u^2}{M_R}$ which, for
$M_R\sim 10^{14}$~GeV, gives $0.1$~eV as needed for resolving
the atmospheric anomaly. As
far as the solar neutrino scale is concerned, introducing an additional
singlet
${\cal N}$ with zero ${\cal U}(1)$ charge, the relevant couplings will be
$F^{(\al )}{\cal N}\ov{\phi }+M_{\cal N}{\cal N}^2$. 
With $M_{\cal N}\sim 3\cdot 10^{15}$~GeV, this gives the desired mass 
$\sim \fr{h_u^2}{M_{\cal N}}\approx 3\cdot 10^{-3}$~eV.

To summarize, an extension of flipped $SU(5)$ GUT by
three vector-like $(F+\ov{F})^{(\al )}$ and some singlet  states 
allows us to exploit the ${\cal U}(1)$ symmetry to generate acceptable
masses and mixings both in the charged fermion and neutrino sectors.

\section{Conclusions}

We have shown that the democratic approach for understanding solar and
atmospheric neutrino oscillations can be nicely implemented within the
MSSM framework and in a suitably extended flipped $SU(5)$ model through
the use of flavor ${\cal U}(1)$ symmetry. It may be possible to extend our
approach to $SO(10)$ which contains flipped $SU(5)$.
In the democratic approach described here the small value of the third
mixing angle $\te_{13}(\stackrel{<}{_\sim }0.2\simeq \ep)$ is due to
accidental cancellations occurring between quantities that have magnitudes
of order unity. In other words, the democratic approach would 
have to be modified if $\te_{13}$ turns out to be much smaller than $\ep$.

\vs{0.2cm} 

\hs{-0.6cm}{\bf Acknowledgments} 

\hs{-0.6cm}Q.S. would like to thank Michael Schmidt and Christof Wetterich
for their
hospitality during his stay at their Institute in Heidelberg, where this
work was initiated. We also acknowledge the support of NATO Grant
PST.CLG.977666. This work is partially supported in part by DOE under
contract DE-FG02-91ER40626.


\bibliographystyle{unsrt}

\begin{thebibliography}{99}


\bibitem{atm} 
%
S. Fukuda et al. [Super-Kamiokande Collaboration],  
Phys. Rev. Lett. 88 (2000) 3999; 
N. Fornengo et al., Nucl. Phys. B 580 (2000) 58. 

\bibitem{sol} 
%
S. Fukuda et al. [Super-Kamiokande Collaboration], 
Phys. Lett. B 539 (2002) 179; 
J. Bahcall, P. Krastev, A. Smirnov, hep-ph/0006078; 
M.C. Gonzalez-Garcia et al., Nucl. Phys. B 573 (2000) 3. 

\bibitem{chooz}
M. Apollonio et al. (CHOOZ Collaboration), Phys. Lett. B 466 (1999) 415.  

\bibitem{feru1} 
%
C.D. Froggatt, H.B. Nielsen, Nucl. Phys. B 147 (1979) 277. 


\bibitem{1feru1}
%
G. Lazarides, Q. Shafi, Nucl. Phys. B 350 (1991) 179;
P. Ramond, R.G. Roberts, G.G. Ross, Nucl. Phys. B 406 (1993) 19;  
L. Iba\~nez, G.G. Ross, Phys. Lett. B 332 (1994) 100; 
P. Binetruy, P. Ramond, Phys. Lett. B 350 (1995) 49; 
V. Jain, R. Shrock, Phys. Lett. B 352 (1995) 83; 
E. Dudas, S. Pokorski, C. Savoy,  
Phys. Lett. B 369 (1995) 255;  
K. Choi, E.J. Chun, H. Kim, Phys. Lett. B 394 (1997) 89; 
Z. Berezhiani, Z. Tavartkiladze,  
Phys. Lett. B 396 (1997) 150; Phys. Lett. B 409 (1997) 220; 
N. Irges, S. Lavignac, P. Ramond, Phys. Rev. D 58 (1998) 035003;
See also references therein.

\bibitem{nuu1}   
P. Binetruy et al., Nucl. Phys. B 496 (1997) 3; 
J. Elwood, N. Irges, P. Ramond, hep-ph/9807228;
J. Sato and T. Yanagida, hep-ph/9809307; 
F. Vissani, hep-ph/9810435;   
Q. Shafi, Z. Tavartkiladze, hep-ph/9807502; hep-ph/9811282;
M. Gomez et al., Phys. Rev. D 59 (1999) 116009; 
C. Froggat, M. Gibson, H. Nielsen, hep-ph/9811265;  
J. Feng, Y. Nir, hep-ph/9911370;  
S. Barr, I. Dorsner, hep-ph/0003058;  
G. Altarelli et al., hep-ph/0007254;  
N. Maekawa, hep-ph/0104200;
I. Gogoladze, A. Perez-Lorenzana, hep-ph/0112034;  
S.F. King, hep-ph/0204360;
T. Ohlsson, G. Seidl, hep-ph/0206087;
See also references therein. 


\bibitem{maxnu1}
Y. Grossman, Y. Nir and Y. Shadmi, hep-ph/9808355; 
Q. Shafi, Z. Tavartkiladze, Phys. Lett. B 448 (1999) 46.

\bibitem{1maxnu1}
S.T. Petcov, Phys. Lett. B 110 (1982) 245; 
R. Barbieri at el., hep-ph/9807235;
A.S. Joshipura, S.D. Rindani, hep-ph/9811252;   
B. Stech, Phys. Lett. B 465 (1999) 219;
R.N. Mohapatra, A. Perez-Lorenzana, C.A. de S. Pires, hep-ph/9911395;
Q. Shafi, Z. Tavartkiladze, Phys. Lett. B 482 (2000) 145; 
T. Kitabayashi, M. Yasue, hep-ph/0006014;  
K.S. Babu, R.N. Mohapatra, hep-ph/0201176.  

\bibitem{2maxnu1} 
Q. Shafi, Z. Tavartkiladze, Phys. Lett. B 451 (1999) 129;
Nucl. Phys. B 573 (2000) 40. 

\bibitem{dem} 
Some authors also use the term {\it neutrino anarchy}.
M. Fukugita, et al., hep-ph/9809554;  
M. Tanimoto, T. Watari, T. Yanagida, hep-ph/9904338;
L. Hall, H. Murayama, N. Weiner, Phys. Rev. Lett. 84 (2000) 2572;
N. Haba, H. Murayama, hep-ph/0009174;
M. Berger, K. Siyeon, hep-ph/0010245;
For neutrino anarchy within a warped 5D scenario see
S. Huber, Q. Shafi, hep-ph/0206258.

\bibitem{1N} 
D. Suematsu, Phys. Lett. B 392 (1997) 413; 
S. Davidson, S.F. King, hep-ph/9808296.

\bibitem{DTsu5}


S. Dimopoulos, F. Wilczek, in Erice Summer Lectures, Plenum, New York,
1981;
H. Georgi, Phys. Lett. B 108 (1982) 283;
B. Grinstein, Nucl. Phys. B 206 (1982) 387;
A. Masiero et al., Phys. Lett. B 115 (1982) 380.


\bibitem{flipsu5}


J.P. Derendinger, J.E. Kim, D.V. Nanopoulos, 
Phys. Lett. B 139 (1984) 170; 
I. Antoniadis et al., 
Phys. Lett. B 194 (1987) 231; Phys. Lett. B 231 (1989) 65;
G.K. Leontaris, J.D. Vergados, Phys. Lett. B 305 (1993) 242;
For the non SUSY case see 
S.M. Barr, Phys. Lett. B 112 (1982) 219;
{}For earlier work on charged fermion masses and neutrino oscillations
within flipped
$SU(5)$ see second ref. in \cite{maxnu1}.



\end{thebibliography}

\end{document}